\documentclass[twocolumn,showpacs,aps,%
prb,amsmath,amssymb,floatfix,superscriptaddress]{revtex4}
\usepackage{amssymb,amsfonts}
\usepackage{epsf,epsfig}
\usepackage{graphicx} 
\usepackage{dcolumn}
\usepackage{bm}

\begin{document}

\title{On the ground state energy scaling in quasi-rung-dimerized spin ladders}
\author{P.~N.~Bibikov}
\affiliation{Saint-Petersburg State University}
\date{\today}

\begin{abstract}
On the basis of periodic boundary conditions we study perturbatively a large $N$ asymptotics ($N$ is the number
of rungs) for the ground state energy density and gas parameter of a spin ladder with slightly destroyed rung-dimerization.
Exactly rung-dimerized spin ladder is treated as the reference model. Explicit perturbative formulas are obtained
for three special classes of spin ladders.
\end{abstract}

\maketitle

\section{Introduction}

Phase structure of frustrated spin ladders and spin ladders with four-spin terms has been intensively studied in the last
decade both theoretically and  numerically \cite{1,2,3}. Among other phases the mathematically most simple one
and at the same time, probably, the one most interesting for physical applications is the so
called rung-singlet (or rung-dimerized) phase \cite{4,5}. Within it the ground state may be well approximated by
an infinite tensor product of  rung-dimers (singlet pairs)
\begin{equation}
|0\rangle_{r-d}=\otimes_n|0\rangle_n.
\end{equation}
This state will be an {\it exact} ground state only for rather big antiferromagnetic rung-coupling and under a special
condition on the coupling constants \cite{4}. The latter has no physical background and thus there are absolutely no
grounds to assume its relevance for real compounds. Nevertheless it is a common opinion that for rather big
antiferromagnetic rung coupling a spin ladder should still remain in the rung-singlet phase. This means that all
physical properties of such a ladder may be obtained perturbatively on the basis of the "bare" ground state (1) and its
excitations. Together with verification by machinery calculations this approach should give a comprehensive description
of the rung-singlet phase. A machinery calculation will provide excellent tests for suggested formulas while a
perturbative formula will give a right direction for numerical research and interpretation of the obtained data.

Such approach has two main difficulties. First of all a general spin ladder model is non-integrable and although
one- and two-magnon states may be readily derived within Bethe Ansatz, three-magnon states are obtained now only
for five special integrable models \cite{6,7}. The second difficulty originates from the fact that an analytical
result is usually obtained for infinite ladder however in a numerical calculation a ladder has a finite size.
Hence in order to use a mashinery calculation for verification of an analytical result one havs to perform a correct
extrapolation of the numerical data. This means that utilizing a finite number of numerical estimations $f_N$ of some
value $f$ ($N$ the number rungs of the ladder) it is nesessary to estimate the limit
$f_{\infty}=\lim_{N\rightarrow\infty}f_N$. On this way, in addition to a number of sequence transformation methods
improving the convergence \cite{8}, one has to be guided by some extrapolation formula.
The latter may be guessed by an analysis of numerical data \cite{9}, or suggested theoreticallly
on the basis of conformal field theory \cite{10}predictions, or on some other argumentation \cite{11}.

Taking an exact rung-dimerized spin ladder as a reference model, it is natural to treat the ground state of a spin
ladder with violated rung-dimerization as a dilute magnon gas \cite{12}. Its consentration (gas parameter)
\begin{equation}
\rho\equiv\rho_{\infty}=\lim_{N\rightarrow\infty}\rho_N,\quad\rho_N=\frac{\langle0|\hat Q|0\rangle}{N},
\end{equation}
($\hat Q$ is a magnon number operator (13)) and energy density
\begin{equation}
E\equiv E_{\infty}=\lim_{N\rightarrow\infty}E_N,\quad E_N=\frac{\langle0|\hat H|0\rangle}{N},
\end{equation}
turns to zero for an exact rung-dimerized spin ladder and hence they should be good governing parameters for a
perturbation theory based on the gas approximation. Perturbative expressions for $\rho$ and $E$ were derived
in Ref. 12. In the present paper assuming {\it periodic} boundary conditions we obtain in three
special cases the corresponding extrapolation formulas for $\rho_N$ and $E_N$.

The two formulas
\begin{eqnarray}
E_N&=&E_{\infty}+(-1)^NA\frac{{\rm e}^{-N/N_0}}{N^2},\\
E_N&=&E_{\infty}-\frac{A}{N^2},
\end{eqnarray}
($A$ and $N_0$ are free parameters) have already been suggested correspondingly for open \cite{13,14} and
periodic \cite{10} boundary conditions. The expression (4) was implied ad hoc, while Eq. (5) follows from conformal
theory argumentation.
The perturbative formulas obtained below for three special classes of spin ladders have a rather different form
\begin{equation}
E_N=E_{\infty}+(A+(-1)^NB){\rm e}^{-(N-1)/N_0}.
\end{equation}

\section{Description of the model}

We shall use an equivalent representation \cite{6,12}
\begin{equation}
\hat H=\hat H_0+J_6\hat V,
\end{equation}
of the spin ladder Hamiltonian \cite{1,2,3,4,5}. Here $J_6$ is a perturbation parameter and
\begin{eqnarray}
\hat H_0&=&\sum_{n=1}^NJ_1Q_n+
J_2({\bf\Psi}_n\cdot{\bf\bar\Psi}_{n+1}+{\bf\bar\Psi}_n\cdot{\bf\Psi}_{n+1})\nonumber\\
&+&J_3Q_nQ_{n+1}+J_4{\bf S}_n\cdot{\bf S}_{n+1}+J_5({\bf
S}_n\cdot{\bf S}_{n+1})^2,\nonumber\\
\hat V&=&\sum_{n=1}^NV_{n,n+1},\\
{\mathbf S}_n&=&{\mathbf S}_{1,n}+{\mathbf S}_{2,n},\quad
Q_n=\frac{1}{2}{\mathbf S}_n^2,\nonumber\\
V_{n,n+1}&=&\tilde{\bf\Psi}_n\cdot\bar{\bf\Psi}_{n+1}+{\bf\Psi}_n\cdot{\bf\Psi}_{n+1},
\end{eqnarray}
(${\bf S}_{i,n}$ for $i=1,2$ are spin-1/2 operators associated
with $n$-th rung). The local operators
\begin{eqnarray}
{\bf\Psi}_n&=&\frac{1}{2}({\bf S}_{1,n}-{\bf S}_{2,n})-i{[}{\bf
S}_{1,n}\times{\bf
S}_{2,n}{]},\nonumber\\
{\bf\bar\Psi}_n&=&\frac{1}{2}({\bf S}_{1,n}-{\bf
S}_{2,n})+i{[}{\bf S}_{1,n}\times{\bf S}_{2,n}{]},
\end{eqnarray}
may be interpreted as (neither Bose nor Fermi)
creation-annihilation operators for rung-triplets. Namely
\begin{eqnarray}
{\bf\bar\Psi}^a_n|0\rangle_n=|1\rangle^a_n,&\quad&{\bf\bar\Psi}^a_n|1\rangle^b_n=0,\nonumber\\
{\bf\Psi}^a_n|0\rangle_n=0,&\quad&
{\bf\Psi}^a_n|1\rangle^b_n=\delta_{ab}|0\rangle_n.
\end{eqnarray}

From (8) and (9) readily follows \cite{6} that
\begin{equation}
[\hat H_0,\hat Q]=0,
\end{equation}
where the operator
\begin{equation}
\hat Q=\sum_nQ_n,
\end{equation}
according to relations
\begin{equation}
Q_n|0\rangle=0,\quad Q_m|1\rangle_n=\delta_{mn}|1\rangle_n,
\end{equation}
has a sence of the number operator for rung-triplets \cite{6}.

For rather big $J_1$ (for example nesessary should be \cite{4,6} $J_1>J_2$) vector (1) is the
{\it zero energy} ($\hat H_0|0\rangle_{r-d}=0$) ground state for $\hat H_0$,
whose physical Hilbert space splits into a direct sum \cite{4,6,12}
\begin{equation}
{\cal H}=\sum_{m=0}^{\infty}{\cal H}^m,\quad \hat Q|_{{\cal
H}^m}=m.
\end{equation}
The subspace ${\cal H}^0$ is generated by the single vector (1). According to (2), (3) and (8)
\begin{equation}
\rho_N=\frac{\partial E_N}{\partial J_1}.
\end{equation}

Since $\hat V:\,|0\rangle_{r-d}\rightarrow{\cal H}^2$, a perturbative treatment of the term $J_6\hat V$
gives
\begin{equation}
E_N=-\frac{J_6^2}{N}\sum_{|\mu\rangle\in{\cal H}^2}\frac{|\langle\mu|\hat
V|0\rangle_{r-d}|^2}{E(\mu)}+o(J_6^2),
\end{equation}
where all the states $|\mu\rangle$ in the sum have zero
total spin and quasimomentum. In the $N\rightarrow\infty$ limit \cite{12}
\begin{eqnarray}
&E_{\infty}=-\Theta(\Delta_0^2-1)\frac{\displaystyle3J_6^2(\Delta_0^2-1)}
{\displaystyle\Delta_0^2E_{bound}}&\nonumber\\
&-\frac{\displaystyle3J_6^2}{\displaystyle4J_2\Delta_0}\left(1-
\frac{\displaystyle J_2|\Delta_0^2-1|+2\Delta_0\sqrt{J_1^2-J_2^2}}{\displaystyle
[2\Delta_0J_1+(\Delta_0^2+1)J_2]}\right),
\end{eqnarray}
where $\Theta(x)=1$ for $x>0$ and $\Theta(x)=0$ for $x\leq0$ and
\begin{eqnarray}
\Delta_0&=&\frac{J_3-2J_4+4J_5}{2J_2},\\
E_{bound}&=&4J_1+2J_2\Big(\Delta_0+\frac{1}{\Delta_0}\Big).
\end{eqnarray}

\section{A finite-$N$ two-particle problem}

A zero total spin and quasimomentum two-magnon state has the
following general form,
\begin{equation}
|2-magn\rangle=\sum_{1\leq m<n\leq N}a(n-m)
...|1\rangle^a_m...|1\rangle^a_n...
\end{equation}
The dimension of the corresponding Hilbert space is $N/2$ for even $N$ and $(N-1)/2$ for odd.
The wave function $a(n)$ should be normalised
\begin{equation}
\sum_{n=1}^{N-1}(N-n)|a(n)|^2=\sum_{m<n}|a(n-m)|^2=\frac{1}{3},
\end{equation}
and satisfy the periodicity condition $a(n-m)=a(m+N-n)$ or shortly
\begin{equation}
a(n)=a(N-n).
\end{equation}
Performing a substitution $n\rightarrow N-n$ and using (23) one can obtain from (22)
\begin{equation}
\sum_{n=1}^{N-1}n|a(N-n)|^2=\sum_{n=1}^{N-1}n|a(n)|^2=\frac{1}{3}.
\end{equation}
Together (22) and (24) result in
\begin{equation}
\sum_{n=1}^{N-1}|a(n)|^2=\frac{2}{3N}.
\end{equation}

The ${\rm Schr\ddot odinger}$ equation gives
\begin{equation}
4J_1a(n)+2J_2{[}a(n-1)+a(n+1){]}=Ea(n)
\end{equation}
for $1<n<N-1$ and
\begin{equation}
2(2J_1+J_2\Delta_0)a(1)+2J_2a(2)=Ea(1),
\end{equation}
for $n=1$.

General solution of the system (26), (27) has the form
\begin{equation}
a(n,z)=\frac{1}{\sqrt{Z(z)}}\left[\Big(1-\frac{\Delta_0}{z}\Big)z^n-\frac{1}{z^n}\Big(1-\Delta_0z\Big)\right],
\end{equation}
and dispersion
\begin{equation}
E(z)=4J_1+2J_2\Big(z+\frac{1}{z}\Big).
\end{equation}
The normalization constant $Z(z)$ ensures condition (25). The parameter $z$ corresponds to relative quasimomentum of
magnon pair and satisfy an equation
\begin{equation}
z^{N-1}=\frac{\Delta_0z-1}{z-\Delta_0}=-z\frac{\Delta_0-1/z}{\Delta_0-z}.
\end{equation}
The latter is invariant under complex conjugation and a duality symmetry
\begin{equation}
z\rightarrow\frac{1}{z},
\end{equation}
which according to (28) is related to multiplication of the wave function on (-1). Hence for even $N$ the roots of (30)
are joined in dual pairs, while for odd $N$ there is an additional autodual root $z=-1$.

In the three special cases $\Delta_0=-1$, $\Delta_0=1$ and $\Delta_0=0$ Eq. (30) may be solved explicitly.
Denoting the corresponding solutions as $u_j$, $v_j$ and $w_j$ respectively one has
\begin{eqnarray}
u_j&=&{\rm e}^{(2j+1)i\pi/(N-1)},\quad j=0,...,N-2,\quad(\Delta_0=-1),\nonumber\\
v_j&=&{\rm e}^{2ji\pi/(N-1)},\quad j=0,...,N-2,\quad(\Delta_0=1),\nonumber\\
w_j&=&{\rm e}^{(2j+1)i\pi/N},\quad j=0,...,N-1,\quad(\Delta_0=0).
\end{eqnarray}

Taking into account that all the roots (32) lie in a unite circle one may readily get
\begin{eqnarray}
Z(z)&=&3N(N-1)(1-\Delta_0z)\Big(1-\frac{\Delta_0}{z}\Big),\quad\Delta_0=\pm1,\nonumber\\
Z(z)&=&3N^2,\quad\Delta=0,
\end{eqnarray}
and then
\begin{eqnarray}
|a(n,z)|^2&=&\frac{1}{3N(N-1)}\Big[2+\Delta_0\Big(z^{2n-1}+\frac{1}{z^{2n-1}}\Big)\Big],\nonumber\\
&&\Delta_0=\pm1,\nonumber\\
|a(n,z)|^2&=&\frac{1}{3N^2}\Big(2-z^{2n}-\frac{1}{z^{2n}}\Big),\quad\Delta_0=0.
\end{eqnarray}

\section{Exact results at $\Delta_0=0$ and $\Delta_0=\pm1$}

Let $|z\rangle$ be the state related to wave function (28).
From (9) and (21) follows that
\begin{eqnarray}
|\langle z|\hat V|0\rangle_{r-d}|^2=9N^2|a(1,z)|^2.
\end{eqnarray}

For the evaluation of $E_N$ one has to perform in (17) a summation over
all duality pairs of roots. Since both the roots in a pair give
the same contribution this is equivalent to inserting the factor
$1/2$ before summation over {\it all} roots. Hence (17) and (35) result in
\begin{equation}
E_N(\Delta_0)=-\frac{3}{4}J_6^2G_N(\Delta_0)+o(J_6^2),
\end{equation}
where
\begin{widetext}
\begin{eqnarray}
G_N(-1)&=&\frac{1}{N-1}\sum_{j=0}^{N-2}\frac{2-(u_j+1/u_j)}{2J_1+J_2(u_j+1/u_j)}=
\frac{1}{J_2(N-1)}\sum_{j=0}^{N-2}\Big[-1+\frac{J_1+J_2}{\sqrt{J_1^2-J_2^2}}\Big(\frac{J_-}{J_--u_j}-
\frac{J_+}{J_+-u_j}\Big)\Big],\nonumber\\
&=&\frac{1}{J_2}\Big[\frac{J_1+J_2}{\sqrt{J_1^2-J_2^2}}\Big(\frac{J_-^{N-1}}{J_-^{N-1}+1}-
\frac{J_+^{N-1}}{J_+^{N-1}+1}\Big)-1\Big],\nonumber\\
G_N(1)&=&\frac{1}{N-1}\sum_{j=0}^{N-2}\frac{\displaystyle2+(v_j+1/v_j)}{2J_1+J_2(v_j+1/v_j)}=
\frac{1}{J_2(N-1)}\sum_{j=0}^{N-2}\Big[1-\frac{J_1-J_2}{\sqrt{J_1^2-J_2^2}}\Big(\frac{J_-}{J_--v_j}-
\frac{J_+}{J_+-v_j}\Big)\Big],\nonumber\\
&=&\frac{1}{J_2}\Big[1-\frac{J_1-J_2}{\sqrt{J_1^2-J_2^2}}\Big(\frac{J_-^{N-1}}{J_-^{N-1}+(-1)^{N-1}}-
\frac{J_+^{N-1}}{J_+^{N-1}+(-1)^{N-1}}\Big)\Big],\nonumber\\
G_N(0)&=&\frac{1}{N}\sum_{j=0}^{N-1}\frac{2w_j^2-w_j^4-1}{w_j(J_2w_j^2-2J_1w_j+J_2)}=
\frac{2}{J_2^2N}\sum_{j=0}^{N-1}\Big[J_1-\frac{J_2}{2}\Big(w_j+\frac{1}{w_j}\Big)
-\sqrt{J_1^2-J_2^2}\Big(\frac{J_-}{J_--w_j}-\frac{J_+}{J_+-w_j}\Big)\Big]\nonumber\\
&=&2\Big[\frac{J_1}{J_2^2}-\frac{\sqrt{J_1^2-J_2^2}}{J_2^2}\Big(\frac{J_-^N}{J_-^N+1}-
\frac{J_+^N}{J_+^N+1}\Big)\Big],
\end{eqnarray}
\end{widetext}
and
\begin{equation}
J_{\pm}=\frac{-J_1\pm\sqrt{J_1^2-J_2^2}}{J_2}.
\end{equation}

In (37) we used for calculations the formulas
\begin{eqnarray}
\sum_{j=0}^{N-2}\frac{1}{J-u_j}&=&\frac{(N-1)J^{N-2}}{J^{N-1}+1},\nonumber\\
\sum_{j=0}^{N-2}\frac{1}{J-v_j}&=&\frac{(N-1)J^{N-2}}{J^{N-1}+(-1)^N},\nonumber\\
\sum_{j=0}^{N-1}\frac{1}{J-w_j}&=&\frac{NJ^{N-1}}{J^N+1},
\end{eqnarray}
which may be proved according to the following argumentation. The sums in (39) are fractions whose numerator and
denominator are symmetric polynomials with respect to $u_j$, $v_j$ and $w_j$ respectively.
However according to (30) all these polynomials exept
\begin{eqnarray}
u_0...u_{N-2}&=&(-1)^{N-1},\quad v_0...v_{N-2}=1,\nonumber\\
w_0...w_{N-1}&=&(-1)^N
\end{eqnarray}
are equal to zero.

From equality $J_+J_-=1$ readily follows
\begin{eqnarray}
\frac{J_-^{N-1}}{J_-^{N-1}+1}-\frac{J_+^{N-1}}{J_+^{N-1}+1}=\frac{1-J_+^{N-1}}{1+J_+^{N-1}},\nonumber\\
\frac{J_-^{N-1}}{J_-^{N-1}+(-1)^{N-1}}-\frac{J_+^{N-1}}{J_+^{N-1}+(-1)^{N-1}}\nonumber\\
=\frac{1-(-J_+)^{N-1}}{1+(-J_+)^{N-1}}.
\end{eqnarray}
Using (41) one may readily reduce Eqs. (37) to the form
\begin{eqnarray}
G_N(-1)&=&\frac{1}{J_2}\left[\sqrt{\frac{J_1+J_2}{J_1-J_2}}\cdot\frac{1-J_+^{N-1}}{1+J_+^{N-1}}-1\right],\nonumber\\
G_N(1)&=&\frac{1}{J_2}\left[1-\sqrt{\frac{J_1-J_2}{J_1+J_2}}\cdot\frac{1-(-J_+)^{N-1}}{1+(-J_+)^{N-1}}\right],\nonumber\\
G_N(0)&=&\frac{2}{J_2}\left[\frac{J_1}{J_2}-\frac{\sqrt{J_1^2-J_2^2}}{J_2}\cdot\frac{1-J_+^N}{1+J_+^N}\right].
\end{eqnarray}

It may be readily observed that the corresponding values for $E_{\infty}(\Delta_0)$ agree with Eq. (18).
The scaling law has the form (6) with
\begin{eqnarray}
A(-1)&=&0,\quad B(-1)=-\frac{3J_6^2}{2J_2}\sqrt{\frac{J_1+J_2}{J_1-J_2}},\nonumber\\
A(1)&=&\frac{3J_6^2}{2J_2}\sqrt{\frac{J_1-J_2}{J_1+J_2}},\quad B(1)=0,\nonumber\\
A(0)&=&0,\quad B(0)=\frac{3J_6^2}{J_2^2}\sqrt{J_1^2-J_2^2},
\end{eqnarray}
at $J_2>0$ and
\begin{eqnarray}
A(-1)&=&-\frac{3J_6^2}{2J_2}\sqrt{\frac{J_1+J_2}{J_1-J_2}},\quad B(-1)=0,\nonumber\\
A(1)&=&0,\quad B(1)=-\frac{3J_6^2}{2J_2}\sqrt{\frac{J_1-J_2}{J_1+J_2}},\nonumber\\
A(0)&=&\frac{3J_6^2}{J_2^2}\sqrt{J_1^2-J_2^2},\quad B(0)=0,
\end{eqnarray}
at $J_2<0$. In both the cases
\begin{equation}
N_0=\frac{1}{\ln{|J_2|}-\ln{(J_1-\sqrt{J_1^2-J_2^2}})}.
\end{equation}

The corresponding formulas for $\rho_N$ have the similar form and may be readily obtained from (16).


\begin{thebibliography}{14}
\bibitem{1} P. Lecheminant, K. Totsuka, Phys. Rev. B {\bf74}, 224426 (2006)
\bibitem{2} M.~T. Batchelor, X.-W. Guan, N. Oelkers, Z. Tsuboi, Adv. Phys. 56, (2007), 465.
\bibitem{3} G. Barcza, $\rm \ddot O$.~Legeza, R.~M. Noack, J.~$\rm S\acute olyom$, Phys. Rev. B {\bf86}, 075133 (2012)
\bibitem{4} A.~K. Kolezhuk, H.-J. Mikeska  {\it Int. J. Mod. Phys.} B {\bf 12}, 2325 (1998)
\bibitem{5} E. Dagotto  Rep. Progr. Phys. {\bf62}, 1525 (1999)
\bibitem{6} P.~N. Bibikov, J. Phys. A {\bf42}, 315212 (2009)
\bibitem{7} P.~N. Bibikov, P.~P. Kulish, Journ. Math. Sci. {\bf168}, 781 (2010)
\bibitem{8} H. Betsuaki, Phys. Rev. B {\bf34}, 8125 (1986)
\bibitem{9} J.~C. Bonner, M.~E. Fisher, Phys. Rev. {\bf135}, A640 (1964)
\bibitem{10} K. Hijii, K. Nomura, Phys. Rev. {\bf65}, 104413 (2002)
\bibitem{11} K. Buchta, $\rm \ddot O$.~Legeza, G. $\rm F\acute ath$, J.~$\rm S\acute olyom$, Phys. Rev. B {\bf72},
054433 (2005)
\bibitem{12} P.~N. Bibikov, Int. Journ. Mod. Phys. B {\bf25}, 1293 (2011)
\bibitem{13} T. Barnes, E. Dagotto, J. Riera, E.~S. Swanson, Phys. Rev. B, {\bf47}, 3196 (1993)
\bibitem{14} D.~B. Yang, W.~C. Haxton, Phys. Rev. B {\bf57}, 10603 (1998)
\end{thebibliography}
\end{document}